\documentclass[aps,pre,preprint,showpacs,superscriptaddress,groupaddress,showkeys,floatfix,byrevtex]{revtex4-1}
\usepackage{graphicx}
\usepackage{dcolumn}
\usepackage{bm}
\usepackage[utf8x]{inputenc}
\usepackage{srctex}
\usepackage{amsmath}
\usepackage{amsfonts}
\usepackage{amssymb}
\usepackage{graphicx}
\usepackage{srctex}
\usepackage{xcolor}
\usepackage{longtable}
\usepackage{subfigure}

\newcommand{\Li}[2]{\hbox{Li}_{#1}\left(#2\right)}
\DeclareMathOperator{\arcsinh}{arcsinh}

\begin{document}
\title{Thermodynamics of the Relativistic Fermi gas in $D$ Dimensions}
\author{Francisco J. Sevilla}
\email[]{fjsevilla@fisica.unam.mx}
\thanks{author to whom correspondence should be addressed.}
\affiliation{Instituto de F\'isica, Universidad Nacional Aut\'onoma de M\'exico, Apdo.\ Postal 20-364, 01000, M\'exico D.F., Mexico}

\author{Omar Pi\~na }
\affiliation{Posgrado en Ciencias F\'isicas, Universidad Nacional Aut\'onoma de M\'exico}
\affiliation{Instituto de F\'isica, Universidad Nacional Aut\'onoma de M\'exico, Apdo.\ Postal 20-364, 01000, M\'exico D.F., Mexico}

\date{Today}

\begin{abstract}
The influence of spatial dimensionality and particle-antiparticle pair production on the thermodynamic properties of the relativistic Fermi gas, at finite chemical potential, is studied. Resembling a kind of phase transition, qualitatively different behaviors of the thermodynamic susceptibilities, namely the isothermal compressibility and the specific heat, are markedly observed at different temperature regimes as function of the system dimensionality and of the rest mass of the particles. A minimum in the isothermal compressibility marks a characteristic temperature, in the range of tenths of the Fermi temperature, at which the system transit from a normal phase, to a phase where the gas compressibility grows as a power law of the temperature. Curiously, we find that for a particle density of a few times the density of nuclear matter, and rest masses $\lesssim 10$ MeV, the minimum of the compressibility occurs at approximately $170\mathrm{MeV}/k_{B},$ which roughly estimates the critical temperature of hot fermions as those occurring in the gluon-quark plasma phase transition.  
\end{abstract}


\maketitle

\section{\label{I} Introduction}

Soon after the discovery of the quantum statistics by Fermi \cite{Fermi1926} and Dirac \cite{Dirac1926}, which incorporates Pauli's exclusion principle \cite{Pauli1925}, the ideal Fermi gas (IFG) has been extensively used to describe, approximately, many physical phenomena in a wide range of values of the particles density, from cosmological scales to nuclear ones. 

At the end of the 20th century, the experimental realization of quantum degeneration of a trapped Fermi gas of $^{40}$K atoms \cite{DemarcoSc} raised the interest on the theoretical study of the thermodynamical and dynamical properties of the Fermi gas in the ideal approximation \cite{Vignolo2000,Gleisber2000,Tran2001,Akdeniz2002,Vignolo2003,Anghel2003,Tran2003,vanZyl2003,Mueller2004,Anghel2005,Song2006}. 

More recently, the IFG has been used in the context of quantum information and the entanglement entropy of it has been obtained in Ref. \cite{GioevPRL2006}, while in Ref. \cite{CalabreseEPL2012} exact relations between the Renyi entanglement entropies and the particle number fluctuations in a system of noninteracting fermions have been derived. 

On the other hand, the thermodynamics of Fermi gases at extreme conditions of density and/or temperature are of great interest to understand processes in white dwarf stars \cite{StockNature1989}, or the properties of the gluon-quark plasma \cite{SatzNature1986}, which is thought to occurred, some microseconds after the Bing-Bang at the early stage of the Universe.  

In the noninteracting regime, analysis of the the relativistic effects on the thermodynamics of the IFG are limited to the consideration of the energy spectrum of a relativistic single-particle \cite{HorePRA75,Dunning-DaviesJPhysA1981,HowardJPhysA2004,ElzeJphysG1980}, generally disregarding particle-antiparticle pair production predicted by quantum field theory. This contrasts to the thermodynamics of the relativistic Bose gas which has been thoroughly studied considering pair production \cite{FrotaPRA1989,GretherPRL2007,SuJPA2008,HaberPRL81}. One of the goals of the present paper is to fill out this gap.

In the same noninteracting regime, P.-H. Chavanis \cite{ChavanisPRE2004,ChavanisPRD2007} discusses the effects of the spatial dimensionality in the balance between quantum pressure due to degeneracy and gravitational collapse due to self-gravitation in white dwarf stars. In his analysis, the author shows that the collapse or evaporation of the star is unavoidable in dimensions larger than four, giving a special character to systems of spatial dimension $d\le3$ through an anthropic principle. The properties of the electron gas in the star are rather well approximated by those in the limit of complete degeneration, {\it i.e.} by those at zero temperature, due to the disparate difference between the system temperature and the Fermi temperature, $T/T_{F}\simeq10^{-3}-10^{-2}.$

Moreover, the effects of low spatial-dimensionality on the non-relativistic IFG at finite temperatures are exhibited in the form of an \emph{unusual} temperature dependence of the chemical potential $\mu(T)$ at constant volume \cite[and reference therein]{GretherEPJD2003}. These effects are markedly shown in an IFG trapped in an impenetrable, one dimensional box potential, for which at low temperatures $\mu(T)$ starts rising quadratically with $T$ above the Fermi energy instead of decreasing from it, as does in the three-dimensional case. Eventually, at larger temperatures, $\mu(T)$ turns to its usual monotonic decreasing behavior at a characteristic temperature that can be as large as twice the Fermi temperature. This unusual behavior has been already related with a possible phase transition with the maximum of the chemical potential pointing out a sort of critical point \cite{sevilla}.

In this paper we focus our study on the thermodynamic properties of the IFG, where the effects of quantum degeneracy, relativity and spatial dimensionality, are all combined. Though, particular attention is paid to the temperature dependence of the chemical potential, which has motivated several discussion of its importance on different levels and contexts \cite{cook_ajp95,baierlein,job2006,MuganEJP2009,ShegelskiSSC86,LandsbergSST87,ShegelskiAJP2004,KaplanJStatPhys2006,SevillaEJP2012}, our main results focus on the thermodynamic susceptibilities or response functions, namely the specific heat at constant volume $C_{V}$ and the isothermal compressibility $\kappa_{T},$ for which there is a great interest at conditions of extreme densities and/or temperatures. 

Interestingly, our calculations reveal the appearance of a transition in the temperature dependence of $C_{V}$ and $\kappa_{T}$ due to pair production, that occurs at a few tenths of the Fermi temperature. This drastic qualitative change in behavior can be plausibly considered as a phase transition, from a phase at which the compressibility diminishes with temperature, as in standard matter (defining the normal phase), to another at which matter becomes arbitrarily compressible. 

The paper is organized as follows. In section \ref{II} we present the system of our study and the chemical potential is calculated from the principle of charge conservation. In section \ref{III} the isothermal compressibility and the heat capacity at constant volume are calculated. Finally, conclusion and final remarks are given in \ref{IV}.

\section{\label{II} Finite temperature: the effects of pair production}

The system under consideration corresponds to a $d$-dimensional gas of non-interacting fermions at finite temperature and chemical potential. We consider the pair production process in thermal equilibrium with a bath of neutral spinless bosons. This corresponds to an over- simplification of the actual physical situation which is under current investigations for interacting systems in Quantum Chromodynamics \cite{Nagata2012} and in the study of collective phenomena \cite{BlaizotPRD2014}. 

At zero temperature the system consists of, without loss of generality, $N_{0}$ of spin-$\frac{1}{2}$ fermions (antifermions may be equally chosen instead), of rest mass $m$ in a volume $V_{d}$, and we consider the exact relativistic energy spectrum, given by
\begin{equation}
E_{k}=\sqrt{c^{2}\hbar ^{2}k^{2}+m^{2}c^{4}},  \label{Energia-Rel}
\end{equation}
where $\hbar k$ is the momentum of the particle and $c$ is the speed of light. For simplicity we assume the spin balanced case in which the number of fermions in each projection $s=\uparrow,\downarrow$ of the spin are equal and spin dependent interactions are neglected.

We introduce the ratio of the rest mass to the \emph{Fermi mass}, $\widetilde{m}=m/m_{F},$ as the parameter that tunes the system from the non-relativistic limit, $\widetilde{m}\gg1$, $E_{k}\simeq mc^{2}+\hbar^{2}k^{2}/2m$, to the ultrarelativistic one $\widetilde{m}\ll1,$ $E_{k}\simeq\hbar c k.$ $m_{F}\equiv\hbar k_{F}/c$ and the Fermi wavevector $k_{F}$ is defined through the Fermi energy $E_{F}\equiv\sqrt{c^{2}\hbar ^{2}k_{F}^{2}+m^{2}c^{4}},$ that gives the energy of the higher occupied state at zero temperature which depends on the zero temperature density of particles in the system $n_{0}=N_{0}/V_{d}$.

In $d$ dimensions the Fermi mass has the following explicit dependence on $n_{0}$
\begin{equation}
m_{F}=2\hbar\pi^{1/2}\left[2\,\Gamma(d/2+1)\right]^{1/d}\, \vert n_{0}\vert^{1/d}/c.
\end{equation}
These relations make clear why, for high dense systems, the ultrarelativistic limit corresponds to $\widetilde{m}\ll1.$  

In the non-relativistc limit $E_{F}\simeq mc^{2}+E_{F}^{NR}$, with $E_{F}^{NR}=\hbar ^{2}k_{F}^{2}/2m$ is the well known non-relativistic Fermi energy. In the opposite limit $\widetilde{m}\ll 1,$ we have $E_{F}=E_{F}^{UR}+\widetilde{m}^{2}/2+\ldots$ with $E_{F}^{UR}=m_{F}c^{2}$. 

According to Quantum Field Theory the relativistic effects of pair production are expected to be important at temperatures of the order of $mc^{2}/k_{B}$ \cite{HaberPRL81,Huang} and the equilibrium state of the mixture of particles-antiparticles is taken into account by the condition $\mu=-\bar{\mu}$ \cite{Huang}, which is straightforwardly obtained by the thermodynamical equilibrium condition on the Helmholtz free energy $F(T,V_{d},N,\overline{N}).$ $N$ and $\overline{N}$ are the number of particle and antiparticles in the system at temperature $T$ and volume $V_{d}$, respectively. Unless otherwise indicated, we denote with an overbar, those quantities related to antiparticles. 

The thermodynamic properties are obtained from the grand partition function 
\begin{equation}\label{PartitionF}
\Xi(T,V_{d},\mu)\equiv\hbox{Tr}\left\{\exp \left({-\beta \left[\boldsymbol{H}-\mu ( \boldsymbol{N}-\overline{\boldsymbol{N}}) \right] }\right)\right\},  
\end{equation}
where $\beta=(k_{B}T)^{-1}$ and $\hbox{Tr}$ denotes the trace over all the states $|n_{\boldsymbol{k}_{1},s}n_{\boldsymbol{k}_{2},s}\ldots\rangle \otimes |\overline{n}_{\boldsymbol{k}_{1},s}\overline{n}_{\boldsymbol{k}_{2},s}\ldots\rangle$ in Fock space, where $\boldsymbol{k}_{i}$ denotes the $d$-dimensional wavevector and $s$ the two projections of spin. $\boldsymbol{H}=\sum_{\boldsymbol{k},s}E_{k}\left(\boldsymbol{n}_{\boldsymbol{k},s}+\overline{\boldsymbol{n}}_{\boldsymbol{k},s}\right),$ $\boldsymbol{N}=\sum_{\boldsymbol{k},s}\boldsymbol{n}_{\boldsymbol{k},s}$ and  $\overline{\boldsymbol{N}}=\sum_{\boldsymbol{k},s}\overline{\boldsymbol{n}}_{\boldsymbol{k},s}$ denote the Hamiltonian, the total number of particles and a anti-particles operators respectively, in terms of the number 
operators $\boldsymbol{n}_{\boldsymbol{k},s}=\boldsymbol{a}^{\dagger}_{\boldsymbol{k},s}\boldsymbol{a}_{\boldsymbol{k},s}$, $\overline{\boldsymbol{n}}_{\boldsymbol{k},s}=\overline{\boldsymbol{a}}^{\dagger}_{\boldsymbol{k},s}\overline{\boldsymbol{a}}_{\boldsymbol{k},s}$, with eigenvalues $n_{k,s}$, $\overline{n}_{k,s}$, where $\boldsymbol{a}^{\dagger}_{\boldsymbol{k},s}$ ($\overline{\boldsymbol{a}}^{\dagger}_{\boldsymbol{k},s}$) and $\boldsymbol{a}_{\boldsymbol{k},s}$ ($\overline{\boldsymbol{a}}_{\boldsymbol{k},s}$) are the creation and annihilation operators of particles (antiparticles) respectively that satisfy the relations of anti-commutation $\left\{\boldsymbol{a}_{\boldsymbol{k}^{\prime},s^{\prime}},\boldsymbol{a}^{\dagger}_{\boldsymbol{k},s}\right\}=\delta_{\boldsymbol{k},\boldsymbol{k}^{\prime}}\delta_{s,s^{\prime}}$, $\left\{\boldsymbol{a}_{\boldsymbol{k}^{\prime},s^{\prime}}^{\dagger},\boldsymbol{a}^{\dagger}_{\boldsymbol{k},s}\right\}=\left\{\boldsymbol{a}_{\boldsymbol{k}^{\prime},s^{\prime}},\
\boldsymbol{a}_{\boldsymbol{k},s}\right\}=0$. The grand canonical partition function results
\begin{equation}
\Xi(T,V_{d},\mu)= \prod_{\boldsymbol{k},s}\left(1+ze^{-\beta E_{k}}\right)\left(1+\overline{z}e^{-\beta E_{k}}\right)
\end{equation}
with $z=e^{\beta\mu},$ $\overline{z}=z^{-1},$ the fugacity of particles and antiparticle respectively. From this, we have that 
\begin{equation}
 \ln\Xi(T,V_{d},\mu)=\sum_{\boldsymbol{k},s}\left[\ln\left(1+ze^{-\beta E_{k}}\right)+\ln\left(1+z^{-1}e^{-\beta E_{k}}\right)\right].
\end{equation}

The net number of particles in the system at $T$ y $V_{d}$ is given by 
\begin{eqnarray}
N-\overline{N}&=& \left[z\frac{\partial\ln\Xi}{\partial z}\right]_{T,V_{d}}\notag \\
&\equiv &\sum_{\boldsymbol{k},s}\left[\langle n_{E_{k}}\rangle-\langle\bar{n}_{E_{k}}\rangle\right],
\end{eqnarray}
where $\langle{n}_{E_{k}}\rangle=\lbrace\exp\left[\beta(E_{k}-\mu)\right]+1\rbrace^{-1}$ and $\langle\bar{n}_{E_k}\rangle=\lbrace\exp \left[ \beta (E_{k}+\mu )\right] +1\rbrace^{-1}$ give, respectively, the average number of fermions and anti-fermions in the energy state $E_{k}.$ 
This equation relates the chemical potential of the system with the initial density of particles $n_{0}$ with $N_{0}=(N-\overline{N})$, in the limit of the continuum we have
\begin{equation}\label{NoEq}
n_{0}=R_{d}\int_{0}^{\infty}dk\, k^{d-1}\left[\langle n_{E_{k}}\rangle-\langle\bar{n}_{E_{k}}\rangle\right],
\end{equation}
where the $R_{d}\equiv4\pi^{d/2}/[(2\pi)^{d}\Gamma(d/2)]$ is a constant that depends only on $d.$ Expression \eqref{NoEq} can be written in terms of hyperbolic functions as
\begin{equation}
n_{0}=R_{d}\int _0^{\infty }dk\, k^{d-1}\frac{\sinh\beta\mu}{\cosh\beta  E_k+\cosh\beta \mu }
\end{equation}
and simplifies to 
\begin{equation}\label{NumEqUR}
n_{0}=-\frac{R_{d}\Gamma(d)}{(\beta\hbar c)^{d}}[\Li{d}{-z}-\Li{d}{-z^{-1}}]
\end{equation}
in the ultrarelativistic limit and to 
\begin{equation}
n_{0}=\frac{R_{d}\Gamma(d/2)}{2(\beta\hbar^{2}/2m)^{d/2}}\left[-\Li{d/2}{-z^{NR}}\right]
\end{equation}
in the non-relativistic one, with $z^{NR}\equiv e^{\beta\mu^{NR}}$ the non-relativistic fugacity and $\mu^{NR}\equiv \mu-mc^{2}$. In last expressions $-\Li{\sigma}{-z}\equiv[1/\Gamma(\sigma)]\int_{0}^{\infty}dx\, x^{\sigma-1}/[e^{x}z^{-1}+1]$ is the polylogarithm function, which has the series representation $-\sum_{l=1}^{\infty}(-z)^{l}/l^{\sigma}$ for $\vert z\vert<1.$

It is worthwhile pointing out that expression \eqref{NumEqUR} can be written as a polynomial of degree $d$ in $(\beta\mu)$ for odd dimension\cite{ElzeJphysG1980}, namely
\begin{equation}
n_{0}=\frac{R_{d}}{d}\left(\frac{\mu}{\hbar c}\right)^{d}\left[1+\sum_{j=0}^{d-2}\frac{d!}{j!}[1+(-1)^{j+d}](1-2^{j+1-d})\zeta(d-j)(\beta\mu)^{j-d}\right]
\end{equation}
where $\zeta(z)$ denotes the usual Riemman zeta function.

\subsection{The chemical potential}

Before discussing the temperature dependence of $\mu$ in the regime of interest, we comment in passing that at low enough temperatures, when pair production is negligible, application of the commonly used Sommerfeld expansion \cite{Ashcroft} to Eq. (\ref{NoEq}), gives for the chemical potential 
\begin{equation}\label{Sommerfeld}
\frac{\mu(T)}{E_F}=1-\frac{\pi^2}{6}\left(\frac{T}{T_F}\right)^{2}\left[1+(d-2)(1+\widetilde{m}^{2})\right],
\end{equation}
which depends explicitly on $\widetilde{m}.$ From this expression, a simple analysis shows that a \emph{non-monotonic} dependence on $T$ is possible whenever the dimensionality of the system is strictly smaller than $2-\left(1+\widetilde{m}^{2}\right)^{-1}$. This result generalizes the one reported in \cite{GretherEPJD2003} by incorporating the finite rest mass effects, and reduces to the inequalities $d<1$ and $d<2$ in the ultrarelativistic and non-relativistic limits respectively. These two extreme values of dimension corresponds to those for which the IFG is thermodynamically equivalent (in that the specific heat has the same temperature dependence) to the ideal Bose gas. 

\begin{figure}[h]
  \centering
   \includegraphics[width=0.7\columnwidth]{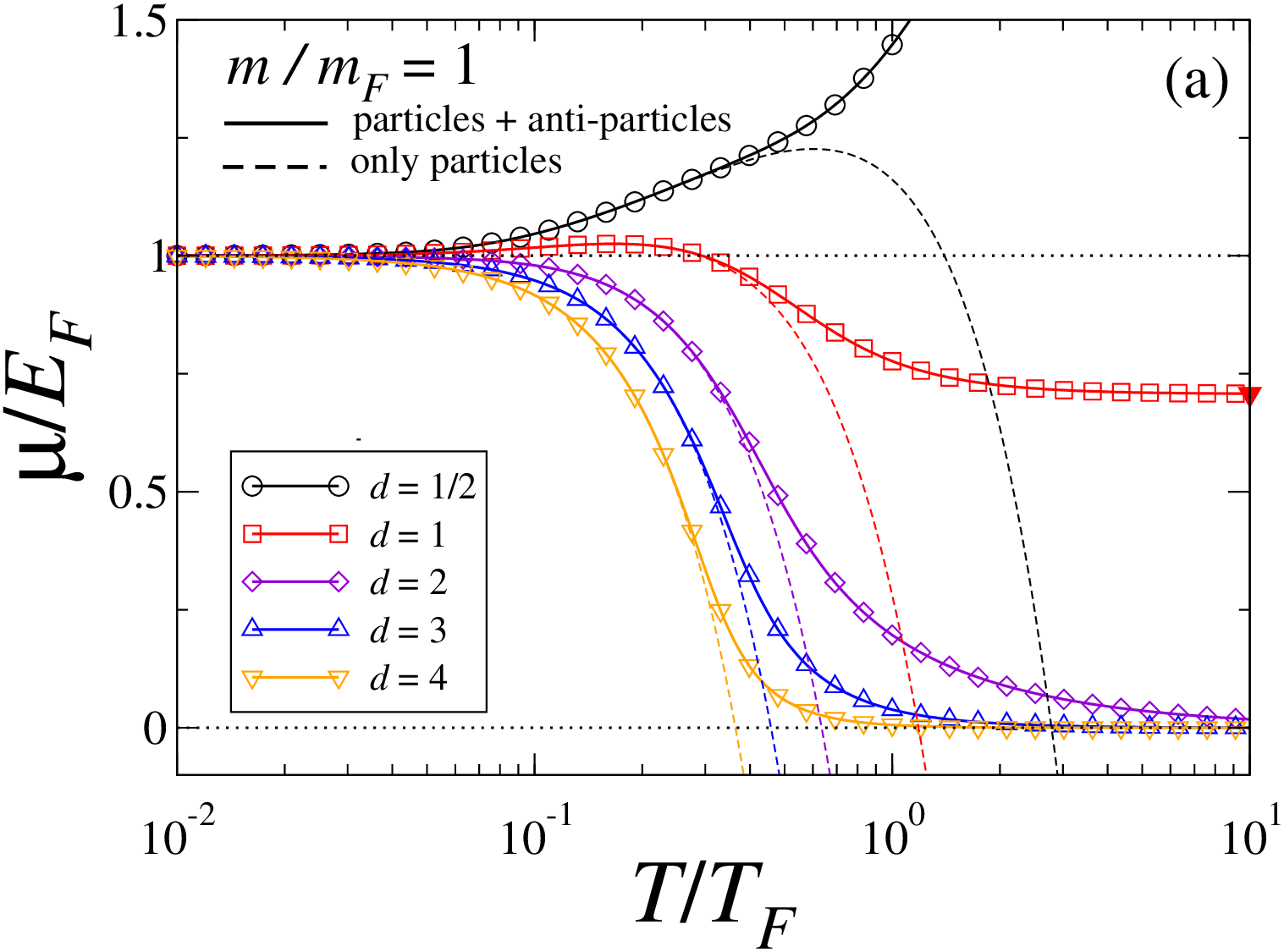}
   \includegraphics[width=0.7\columnwidth]{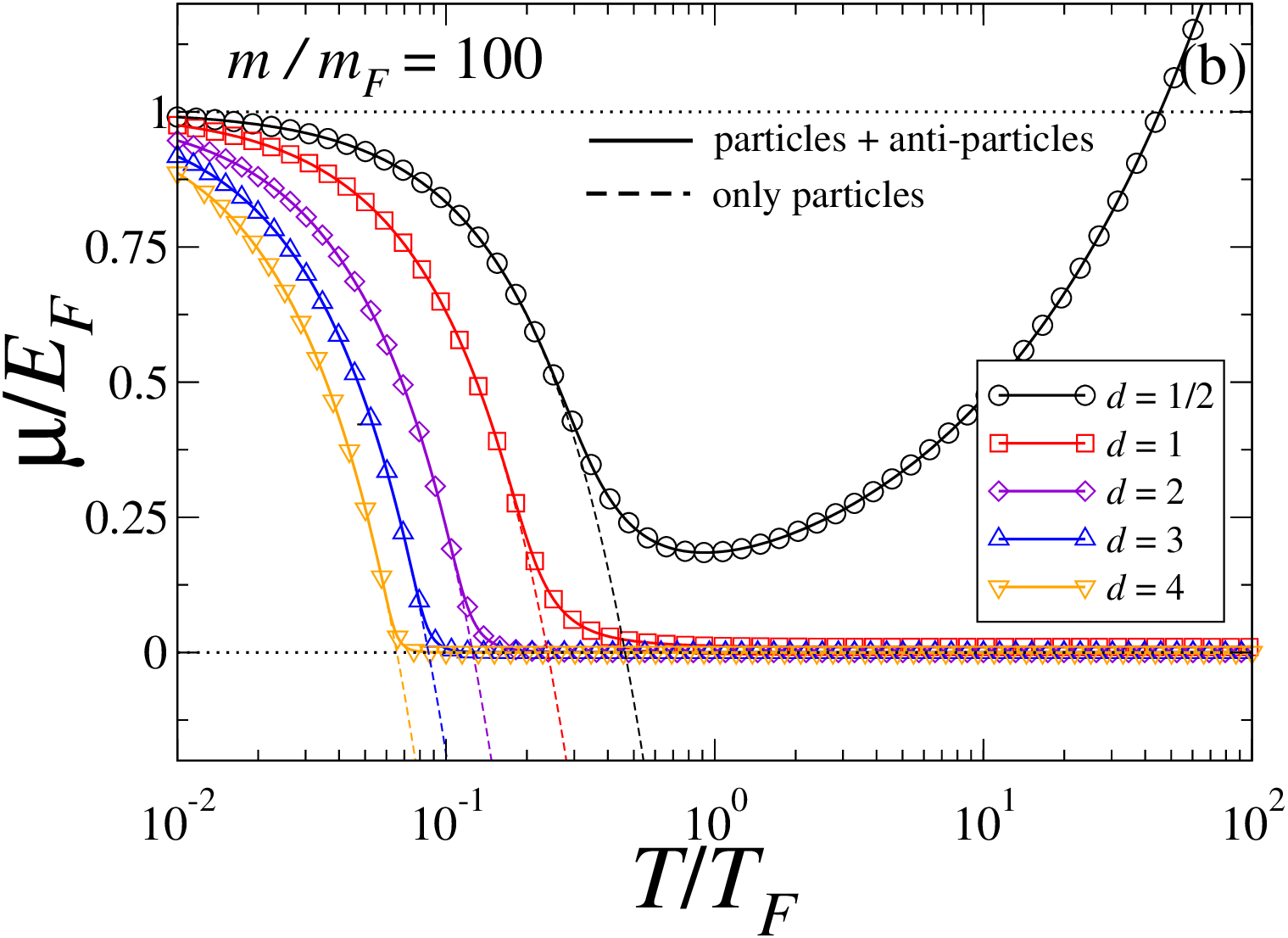}
  \caption{(Color online) Dimensionless chemical potential as function of the dimensionless temperature $T/T_{F}$, for different dimensions: 1/2 (circles), 1 (squares), 2 (diamonds), 3 (up-triangles) and 4 (down-triangles). Panel (a), (b), corresponds to $\widetilde{m}=\,1,$ 100, respectively. To exhibit the effects of pair production, the curves of $\mu(T,V_{d})$ for the case when only particles are present in the system are also shown (dashed lines). The solid-red triangle in panel (a) gives the value $1/\sqrt{2}$ which corresponds to the asymptotic value given by expression \eqref{muasymptotic} for $d=1$.}
\label{fig:ChemPot}
\end{figure}

In Fig.\ref{fig:ChemPot}(a) $\mu(T)$ without pair production  is shown (dashed lines) for dimensions 1/2, 1 ,2, 3 and 4 for $\widetilde{m}=1$. The non-monotonous behavior expected for $d<2$, is exhibited as a local maximum for dimensions 1/2 (first dashed line from the far right) and 1 (second dashed line from the far right). These maxima survive only for $d<1$ in the limit $\widetilde{m}\rightarrow0.$ For $\widetilde{m}=100$ Fig.\ref{fig:ChemPot}(b), the chemical potential apparently does not reveal a maximum value expected for $d<2,$ this is because we have chosen to scale $\mu$ with the exact $E_{F},$ however by shiftting by $mc^{2}$ and changing the scaling factor to $E_{F}^{NR}$  we recover the non-monotonic behavior of   the non-relativistic IFG $\mu^{NR}=\mu-mc^{2}$ (see Fig.$1$ in Ref.\cite{GretherEPJD2003}).

Without pair production and in the high temperature regime $\mu$ is given by
\begin{equation*}
-k_{B}T\, \ln\left[\frac{V_{d}}{N\lambda^{d}}\, 2\left(\frac{2\pi k_{B}T}{mc^{2}}\right)^{(d-1)/2} K_{(d+1)/2}\left(\frac{mc^{2}}{k_{B}T}\right)\right],
\end{equation*}
where $\lambda=h/mc$ is the Compton wavelength and $K_{\nu}(z)$ denotes the Bessel function of the second kind of order $\nu.$ Last expression corresponds to the classical result for which the chemical potential is negative and decreases monotonically with temperature (see dashed lines in Fig. \ref{fig:ChemPot}). In addition, the same expression is also obtained for $N$ spin-less relativistic bosons of mass $m$ in the same limit \cite{FrotaPRA1989}. This trivial relationship between the Bose and Fermi gas is simply established by the loss of quantum degeneracy due to thermal fluctuations. 

\paragraph{Effects of pair production.}
By solving Eq. (\ref{NoEq}) at constant volume, we show that the combined effects of pair production and system dimensionality are conspicuous on the temperature dependence of $\mu(V_{d},T)$ as is shown in Fig. \ref{fig:ChemPot} (solid lines with symbols).

In the high temperature regime, the chemical potential has three distinct asymptotic limits: i) it goes to zero if $d>1$; ii) it goes to the constant value $E_{F}\left[1+\widetilde{m}^{2}\right]^{-1/2}$ if $d=1$ and iii) diverge sub-linearly as a power law for $0<d<1.$ These behaviors are accounted for by the expression
\begin{equation}\label{muasymptotic}
\mu(T)\sim E_{F}\left(\frac{T}{T_{F}}\right)^{1-d}\Phi(\widetilde{m}^{2}+1,d), 
\end{equation}
which is approximately obtained from Eq. (\ref{NoEq}), with $\Phi(\xi,d)$ a temperature-independent quantity defined through the expression
\begin{equation}\label{Phi}
\left[\Phi(\xi,d)\right]^{-1}=d\int_{0}^{\infty}dx\, x^{d-1}\left[1+\cosh\left(\frac{x^{2}}{\xi}\right)^{1/2}\right]^{-1}.
\end{equation}
In Table \ref{table} explicit functional forms for $\left[\Phi(\xi,d)\right]^{-1}$ are given for $d=4,\,3,\,2$ and 1. 
\begin{table}[h]
\caption{Explicit functional forms for $\left[\Phi(\xi,d)\right]^{-1}$ which appears in eq. \eqref{muasymptotic}.}
\begin{tabular}{|c|c|c|c|c|}
\hline
 & $d=4$ & $d=3$ & $d=2$ & $d=1$ \\ \hline
$\Phi(\xi,d)^{-1}$ & $36\, \xi^2\, \zeta(3)$ & $\pi^{2}\, \xi^{3/2}$ & $2\, \xi\,\ln 4$  & $\xi^{1/2}$\\ \hline
\end{tabular}
\label{table}
\end{table}

In the ultrarelativistic limit, the explicit dependence on temperature can be obtained for odd dimensions, namely $\mu/E_{F}=1$ for dimension one and
\begin{equation}
 \frac{\mu}{E_{F}}=\left[\frac{1}{2}+\sqrt{\left(\frac{\pi}{\sqrt{3}}\frac{T}{T_{F}}\right)^{6}+\frac{1}{4}}\right]^{1/3}+\left[\frac{1}{2}-\sqrt{\left(\frac{\pi}{\sqrt{3}}\frac{T}{T_{F}}\right)^{6}+\frac{1}{4}}\right]^{1/3}
\end{equation}
for the three-dimensional case \cite{ElzeJphysG1980}.

In Fig.\ref{fig:ChemPot} the effects of pair production on $\mu(T)$ are shown (solid-lines with different symbols which denote different values of the system dimensionality), for the mass ratio $\widetilde{m}=1$ (panel (a)) and $\widetilde{m}=100$ (panel (b)). In both cases, the solid-red line with squares which corresponds to $d=1$, marks the division from the two different behaviors i) and iii).

The behavior exhibited for $d<1$ is puzzling. Though, thermodynamics at these dimensions would seem out of place, the limit $d\rightarrow0$ has been analyzed in Ref. \cite{LeePRE1996} for the non-relativistic IFG, giving a physically consistent interpretation on the meaning of the large values of the chemical potential as $d\rightarrow0$ \cite{ChavezPhysicaE2011}. The effects of pair production makes the chemical potential to grow monotonically for all temperature if $d<1$ and $\widetilde{m}\lesssim2$, since the larger the temperature the larger the number of particles in the system, thus $\langle n_{E_{k}}\rangle>1/2$ for all $T,$ instead of diminishing as happens when the chemical potential decreases monotonically. This considerations make clear why the system at high temperature behaves quite differently from the classical gas counterpart for which $\langle n_{E_{k}}\rangle\ll1$. This explanation agrees with the one given in Ref. \cite{LeePRE1996} for the non-relativistic IFG in diminishing dimensions, where it is argued that the high values of $\mu$ are a manifestation of the Pauli exclusion principle. For larger masses, $\widetilde{m}\gtrsim4$, this behavior is changed as can be noticed in Fig.\ref{fig:ChemPot}(b) for $\widetilde{m}=100$, where the chemical potential exhibits a different non-monotonic behavior, it goes from a decreasing behavior to an increasing one. 

\begin{figure}[ht!]
  \centering
    \includegraphics[width=0.7\columnwidth]{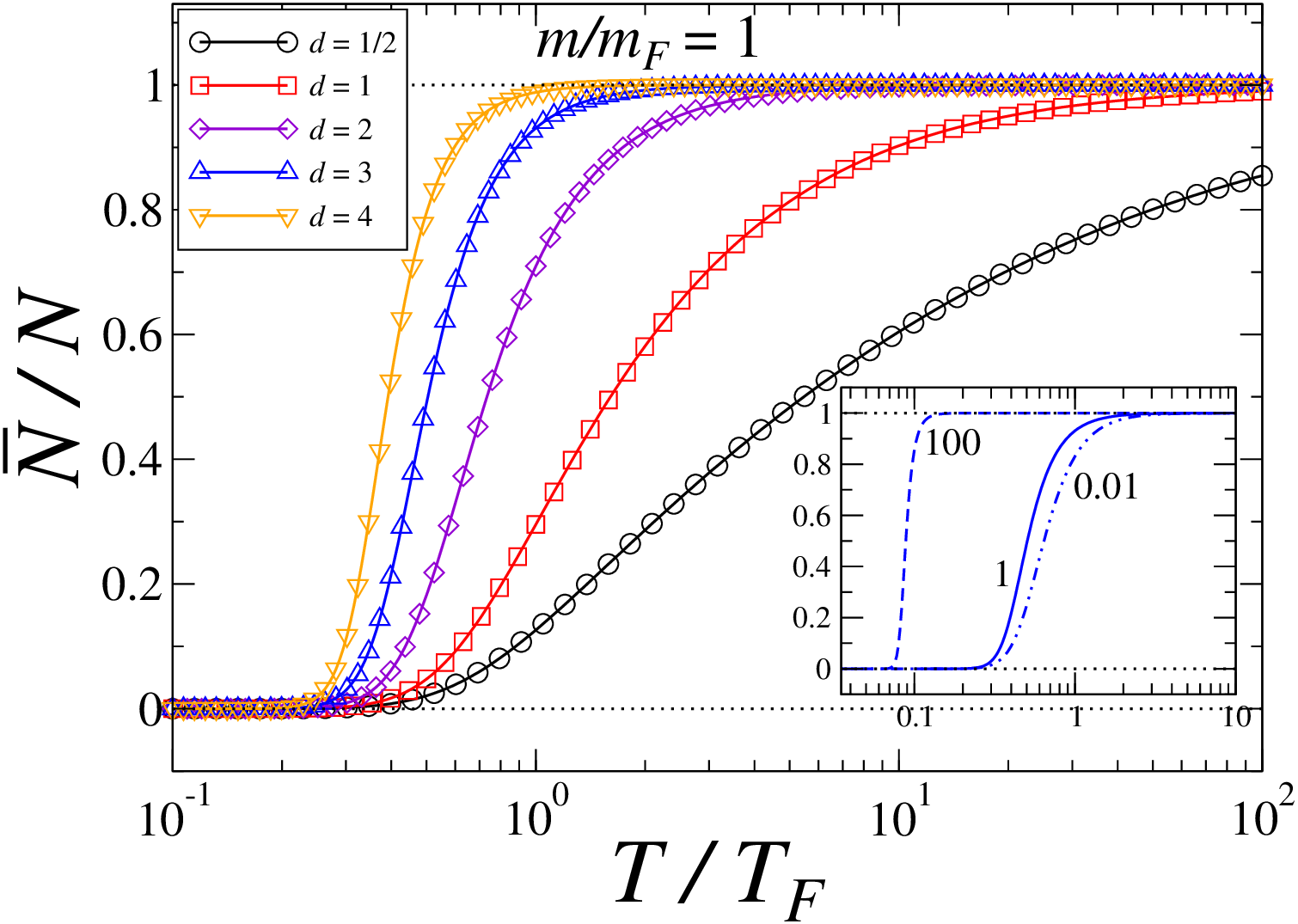}
  \caption[]{(Color online) Ratio of anti-particles number to the particle number as function of $T/T_{F}$ for dimension 1/2 (circles), 1 (squares), 2 (diamonds), 3 (up-triangles) and 4 (down-triangles). Inset correspond to the three-dimensional case for the mass ratio values $\widetilde{m}=0.01$ (dashed-dotted line), $1$ (continuous line) and $100$ (dashed line).}
\label{fig:ratio}
\end{figure}

For $d>1$, the effects of the original number of particles are outweighed by the large rate of pair production, thus tending to the limit $N\approx\overline{N}$. This is so due to the  dependence on $T$ of $\mu,$ which goes to zero as $T^{1-d}$, this implies $z\rightarrow\overline{z}$.  
The particular dependence of $\mu$ on $T,$ for different dimensions and values of $\widetilde{m},$ leads to different particle-antiparticle pair production rate as is exhibited in Fig.\ref{fig:ratio}, where the ratio of the antiparticles number to the particles number, $\overline{N}/N$, is shown as function of temperature for $\widetilde{m}=1$. In the inset, the effects of disparate masses, namely $\widetilde{m}=0.01,\, 1,\, 100,$ are shown for $d=3.$ 

\section{\label{III} Thermodynamic susceptibilities}

It has been suggested \cite{sevilla}, on the grounds of the energy-entropy argument that the non-monotonic behavior of the chemical potential can be related to a phase transition. Indeed, for temperatures below the temperature at which $\mu$ acquires a maximum value, the change of the system's free energy is dominated by the change in the internal energy, while for temperatures above such temperature, the free energy change is dominated by entropic differences. Possible phase transitions are pointed out by the thermodynamical susceptibilities, in particular by the isothermal compressibility $\kappa_{T},$ which has been directly measured for a neutral interacting-fermionic gas and revealed a clear signature of the superfluid transition \cite{KuScience2012}.   

The susceptibilities also play an important role in equilibrium transformations, such as the cooling by adiabatic compression or by an isocoric transformation of a gas.  In such cases, the constant volume specific heat $C_{V}$ and the isothermal compressibility $\kappa_{T}$ are of particular importance. 

Regarding the $C_{V},$ it is known that the non-relativistic IFG shows a monotonic non-decreasing behavior as function of $T$ for dimensions $d\ge2$ and a ``hump'' is developed for $d<2$ \cite{GretherEPJD2003}. The hump is related to the non-monotonic behavior of $\mu(T)$ and means that, for low dimensional systems, the IFG dissipate thermal fluctuations more effectively in the temperature region where $\mu>E_{F}$. The isothermal compressibility also exhibits a ``hump'' for $d<2$ with a maximum at a characteristic temperature $T_{\kappa}$ \cite{SevillaPina}. For $T>T_{\kappa}$ the system compres\-sibility diminish, vanishing as the temperature goes to infinity just like the ideal classical gas. Unexpectedly, below $T_{\kappa}$, the compressibility of the gas rises with $T$ above its $\kappa_{0}$ value, the gas   turns to be more compresible than the $T=0$ state. This qualitative change in the behavior of the IFG in low dimensions has been suggested to be related to a phase transition \cite{sevilla, SevillaPina}. In addition,  a thermodynamic ``equivalence'' between the ideal Bose and Fermi gases in $d=2$ has been analyzed \cite{MayPR64,LeePRE97} and extended to a more general energy-momentum dispersion relation \cite{PathriaPRE98}. Such equivalence is understood as the fact that both gases have the same temperature dependence of their respective specific heat at constant volume. 

\begin{figure}
 \includegraphics[width=0.7\columnwidth]{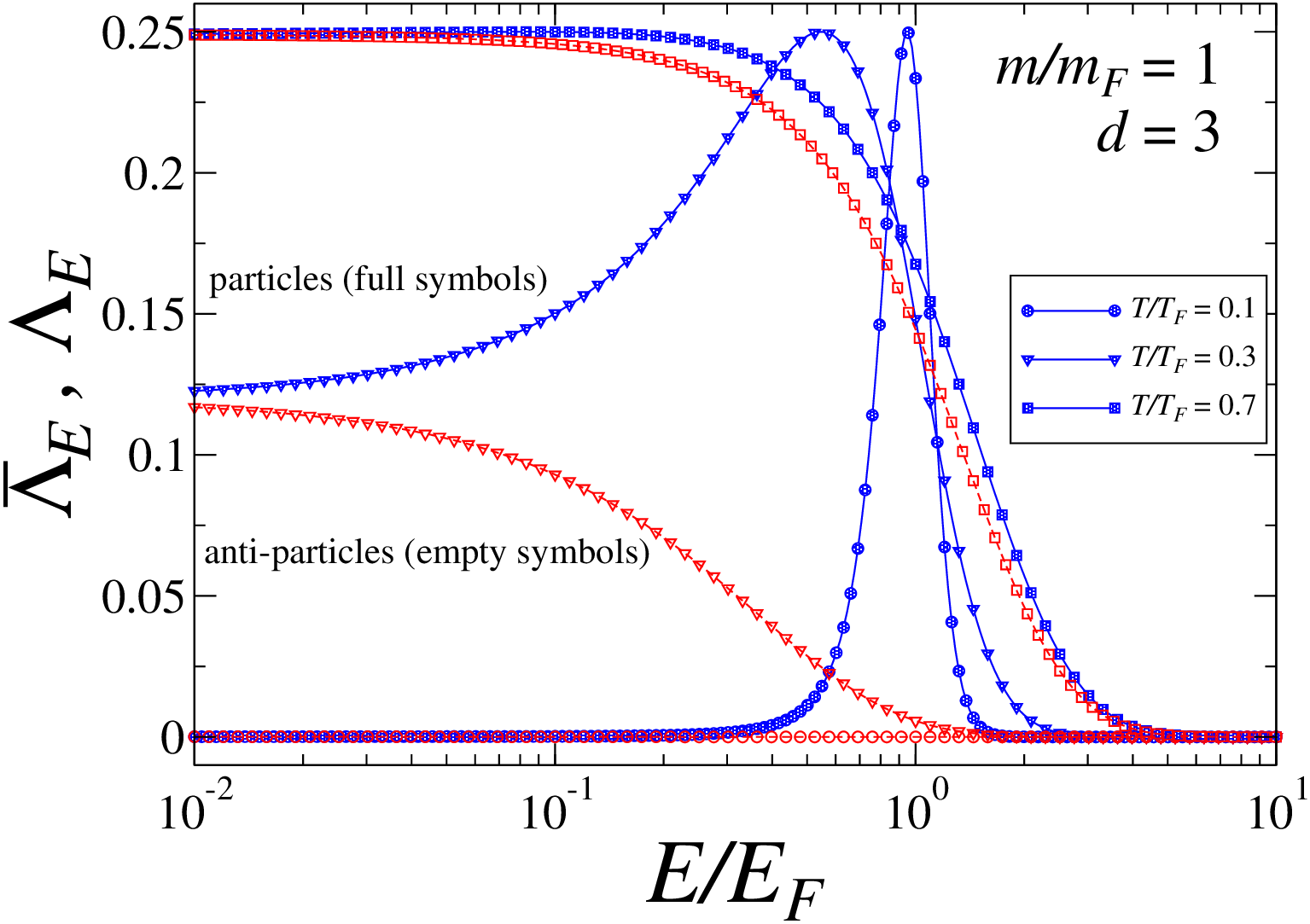}
 \caption{(Color online) $\Lambda_{E},$ $\overline{\Lambda}_{E}$ (as defined in text) \emph{vs} the normalized energy $E/E_{F}$ are shown for the three-dimensional relativistic IFG and for different temperatures, namely $T/T_{F}=0.1$ at which pair production is negligible (circles); $T/T_{F}=0.3$ when pair production starts rising (triangles); and $T/T_{F}=0.7$ where antiparticles almost equals the particles number (squares).}
 \label{lambda_fluctuations}
\end{figure}

Now we turn to analyze the effects of pair production on $\kappa_{T}$ and $C_{V}$ of the relativistic IFG, A quantity of interest that is relevant in the study of the system thermodynamic fluctuations corresponds to $\langle{n}_{E_{k}}\rangle\left(1-\langle{n}_{E_{k}}\rangle\right),$ denoted with $\Lambda_{E_{k}},$ which gives account of the variance of the occupation number of the energy-state  $E_{k}.$  The dependence on $\beta$ and $\mu$ has not been made explicit for the economy of writing. In figure \ref{lambda_fluctuations} we present $\Lambda_{E}$ and $\overline{\Lambda}_{E}$ for the three-dimensional case as function of $E,$ for $\widetilde{m}=1$ and for temperatures at which: a) pair production is negligible $T/T_{F}=0.1$ (circles); b) pair production starts rising $T/T_{F}=0.3$ (triangles); and c) $T/T_{F}=0.7$ where antiparticles almost equals the particles number (squares).     

\subsection{The isothermal compressibility $\kappa_{T}$}
The isothermal compressibility is worth of analysis since is directly related to the number fluctuations of the system and such quantity can be used to characterize many situations of the IFG, as entanglement of the system \cite{CalabreseEPL2012} for instance.

At finite temperature, $\kappa_{T}$ can be computed from the expression $(1/n_{0}^{2})\left(\partial n_{0}/\partial\mu\right)_{T}$ which results into
\begin{equation}
\kappa_{T}=\frac{R_{d}}{n_{0}^{2}k_{B}T}\int_{0}^{\infty}dk\, k^{d-1}\left[\Lambda_{E_{k}}+\overline{\Lambda}_{E_{k}}\right].
\end{equation}

In the ultrarelativistic limit last expression simplifies, in terms of polylogarithms, to 
\begin{equation}\label{CompressUR}
\kappa_{T}=\frac{R_{d}\Gamma(d)}{n_{0}^{2}(\hbar c)^{d}}(k_{B}T)^{d-1}\left[-\hbox{Li}_{d-1}(-z)-\hbox{Li}_{d-1}(-z^{-1})\right]
\end{equation}
and to
\begin{equation}
\kappa_{T}=-\frac{R_{d}\Gamma(d/2)}{n_{0}^{2}\left(\hbar^{2}/2m\right)^{d/2-1}}(k_{B}T)^{d/2-1}\hbox{Li}_{d/2-1}(-z^{NR})
\end{equation}
in the non-relativistic one.

Expression \eqref{CompressUR} can be written as an even polynomial of order $d-1$ in $\beta\mu$ since for odd dimension, the term within square parenthesis can be written as
\begin{equation}\label{kapPolyn}
(-1)^{(d-1)/2}\frac{(2\pi)^{d-1}}{(d-1)!}\sum_{l=0}^{(d-1)/2}(-1)^{l}\eta_{l,d}\frac{(\beta\mu)^{2l}}{(2\pi)^{2l}} 
\end{equation}
where the coefficients $\eta_{l,n}$ are given in terms of the Bernoulli numbers $B_{j}.$ Some of the coefficients are: $\eta_{(d-1)/2,d}=B_{0},$ \ldots, $\eta_{1,d}=\binom{d-1}{d-3}\sum_{j=0}^{d-3}\binom{2+j}{2}\frac{B_{j}}{2^{j}}$, $\eta_{0,d}=\sum_{j=0}^{d-1}\binom{d-1}{d-j-1}\frac{B_{j}}{2^{d-j-1}}.$ Explicit expressions for expression \eqref{kapPolyn} are given in Table \ref{Tab2} for dimensions 1, 2, and 3.
\begin{table}[h]
\caption{Some explicit even polynomial in $\beta\mu$ calculated from expression \eqref{kapPolyn}.}
 \begin{tabular}{|c|c|}
 \hline
  dimension $d$ & $-\hbox{Li}_{d-1}(-z)-\hbox{Li}_{d-1}(-z^{-1})$ \\
  \hline
  1 & 1 \\

  3 & $\zeta(2)+\dfrac{1}{2}(\beta\mu)^{2}$\\

  5 & $\dfrac{7}{4}\zeta(4)+\dfrac{1}{2}\zeta(2)(\beta\mu)^{2}+\dfrac{1}{4!}(\beta\mu)^{4}$\\
  \hline
 \end{tabular}
 \label{Tab2}
\end{table}

\begin{figure}[ht!]
  \centering
   \includegraphics[width=0.7\columnwidth]{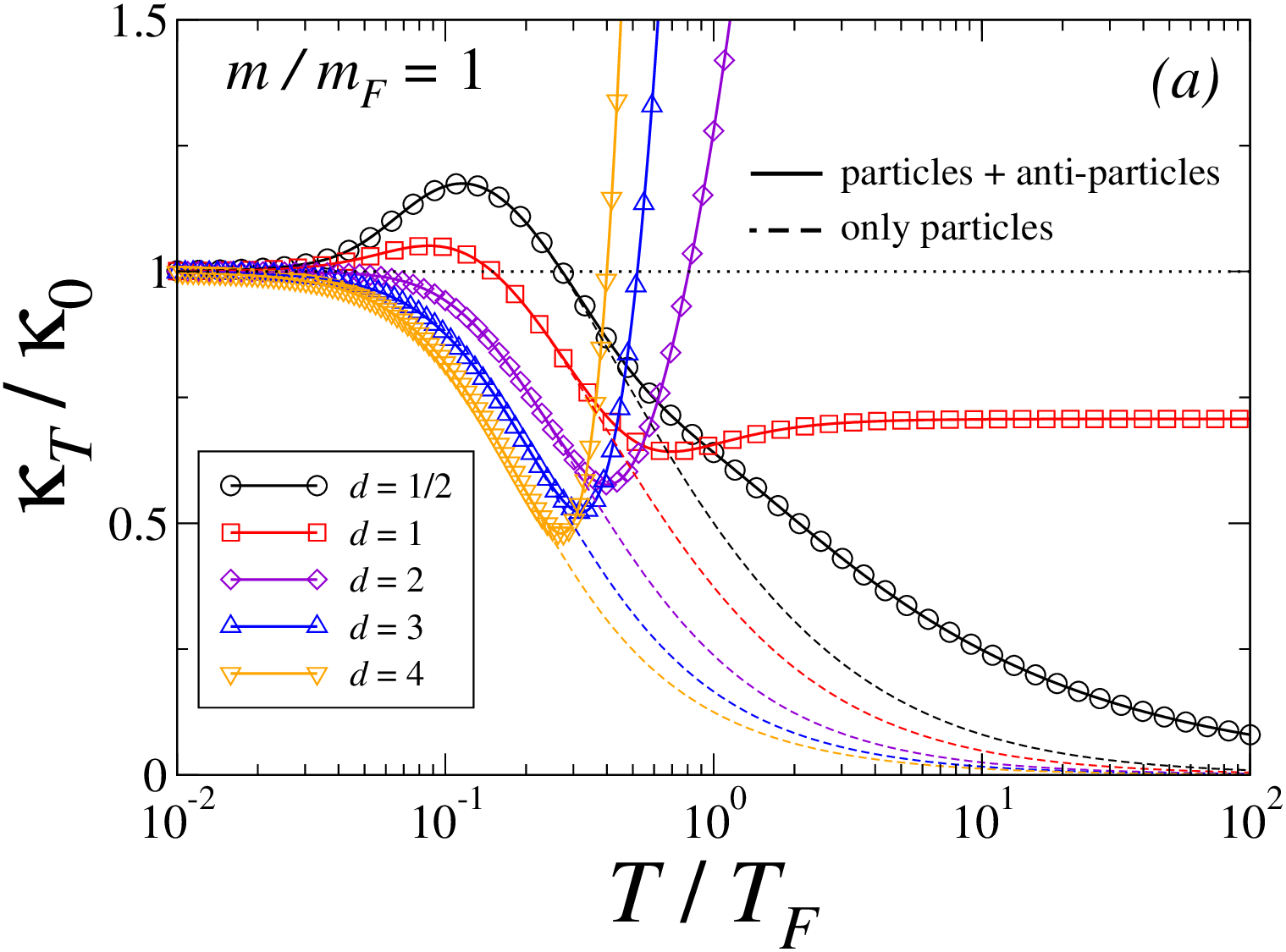}
   \includegraphics[width=0.7\columnwidth]{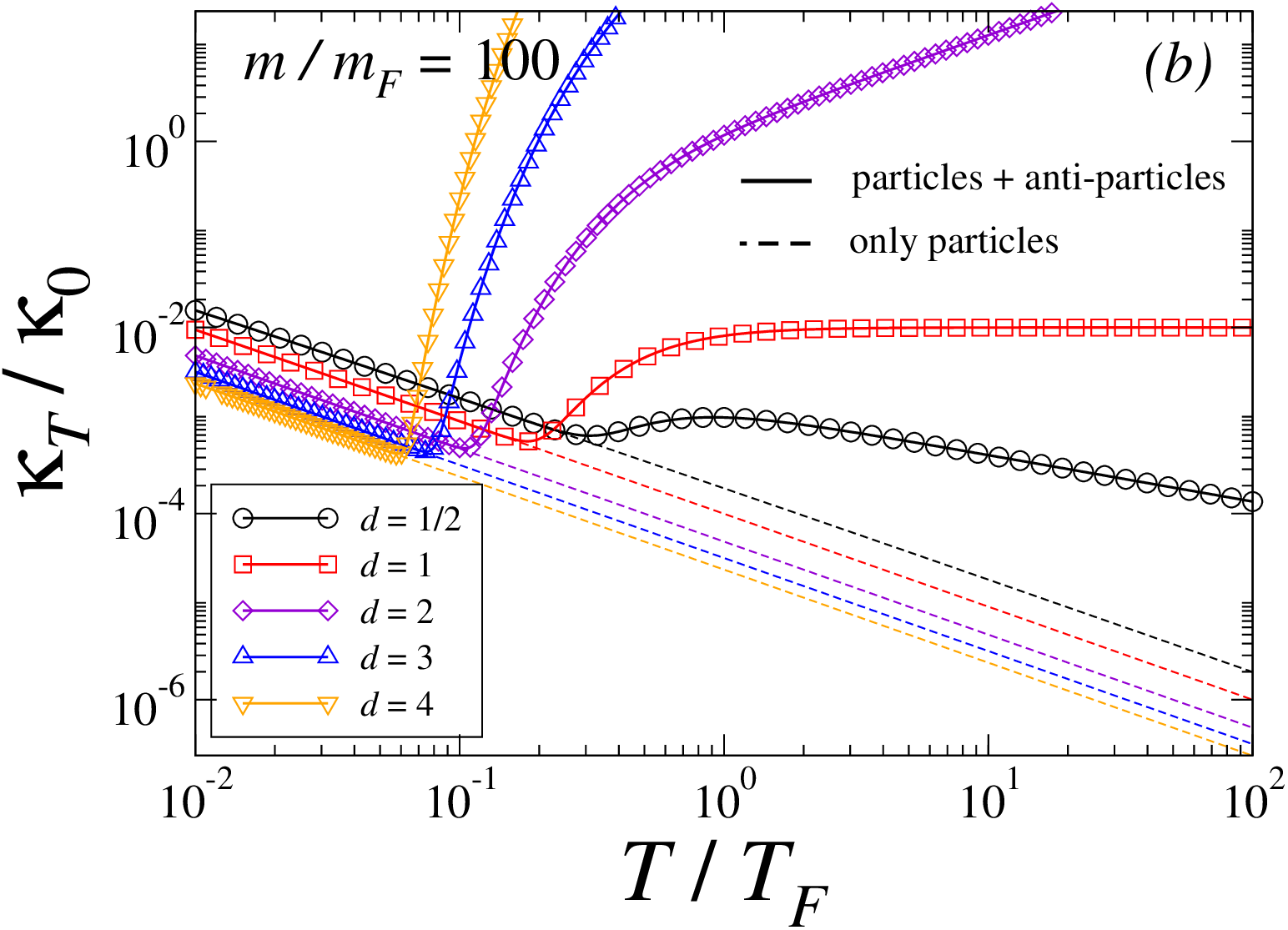}
  \caption{(Color online) Isothermal compressibility $\kappa_{T}$ normalized with its value a $T=0$ as function of the dimensionless temperature $T/T_{F}$ and dimension 1/2 (circles), 1 (squares), 2 (diamonds), 3 (up-triangles) and 4 (down-triangles) for the mass ratio $\widetilde{m}=1$ [panel (a)] and $\widetilde{m}=100$ [panel (b)]. Thin-dashed lines correspond to the cases for which pair production is neglected.}
\label{fig:compressibility}
\end{figure}

In Fig. \ref{fig:compressibility}, $\kappa_{T}$ is shown as function of temperature for $\widetilde{m}=1$ [relativistic case, panel (a)] and $\widetilde{m}=100$ [non-relativistic case, panel (b)] and $d=1/2,\,1,\,2,\,3,$ and $4$. The ultrarelativistic limit $\widetilde{m}\ll1$, has been omitted since analytical expression have been obtained. 

In the low temperature regime, the compressibility rises and eventually starts diminishing with temperature exhibiting a maximum at $T_{\kappa}$. This behavior is determined by $\widetilde{m}$ and $d$. A calculation based on the observation that the product $\langle n_{E}\rangle(1-\langle n_{E}\rangle)$ is different from zero only in a narrow interval of energies around $\mu,$ gives up to second order terms in $T/T_{F}$
\begin{equation}
\kappa_{T}\simeq\frac{d \mu\left(\mu^{2}-m^{2}c^{4}\right)^{d/2-1}}{n_{0}(m_{F}c^{2})^{d}}\left(1+\frac{\pi^{2}}{6}\left(k_{B}T\right)^{2}(d-2)\left(\mu^{2}-m^{2}c^{4}\right)^{-2}\left[3\left(\mu^{2}-m^{2}c^{4}\right)+(d-4)\mu^{2}\right]\right)
\end{equation}
and by using Eq. (\ref{Sommerfeld}) we have that 
\begin{equation}
\frac{\kappa_{T}}{\kappa_{0}}\simeq1-\frac{\pi^{2}}{6}\left(\frac{T}{T_{F}}\right)^{2}\left[1-2(d-2)(1+\widetilde{m}^{2})-(d-2)(d-4)(1+\widetilde{m}^{2})^{2}\right],
\end{equation}
which shows a nonmonotonic dependence with temperature whenever
\begin{equation}\label{conditiondim2}
d<\frac{2+3\widetilde{m}^{2}-\left(1+\widetilde{m}^{4}\right)^{1/2}}{(1+\widetilde{m}^{2})}.
\end{equation}
This raising of the compressibility with temperature is an abnormal feature that would have important effects on some thermodynamical transformations in low dimensional systems at low temperatures \cite{SayginJApplPhys2001,SismanApplEn2001}.

In the limit $\widetilde{m}\rightarrow0$ such abnormal behavior is presented for systems in dimensions smaller than 1, as can be checked from the expression
\begin{equation}
\frac{\kappa_{T}}{\kappa_{0}}=1+\frac{\pi^{2}}{6}(d-1)(d-2)\left(\frac{T}{T_{F}}\right)^{2}
\end{equation}
or directly from \eqref{conditiondim2}.

As can be checked straightforwardly from condition \eqref{CompresibilityUR} $\kappa_{T}$ becomes temperature independent for $d=1$ getting the value $\kappa_{0}^{UR}=\left(\pi\hbar c\, n_{0}^{2}\right)^{-1}$ [see  Eq. (\ref{CompresibilityUR}) in the Appendix]. Note that for the case $d=2$ $\kappa_{T}$ is proportional to $\mu,$ which turns to be a monotonic decreasing function of $T$ for any value of $\widetilde{m}$. 

For $\widetilde{m}\gg1,$ condition \eqref{conditiondim2} turns into $d<2$ which corresponds to the case analyzed by Sevilla and Pi\~na \cite{SevillaPina}. 

As temperature is increased, $\kappa_{T}$ suffers a striking change for $d>1$ at the temperature when pair creation starts to be important. Instead of diminishing to zero as occurs if pair creation is neglected (dashed lines in Fig.\ref{fig:compressibility}), it starts to grow with temperature. This behavior is set on when the number of antiparticles is of the order of particles and is marked by a local minimum $T_{\kappa}^{*}$ in the range of tenths of the Fermi temperature. At higher temperatures, $\kappa_{T}$ grows with $T$ asymptotically as 
\begin{equation*}
\kappa_{0}\left(1+\widetilde{m}^{2}\right)^{d/2-1}\, (d-1)!\, \zeta(d-1)\,2\left(1-2^{2-d}\right)\left(\frac{T}{T_{F}}\right)^{d-1},
\end{equation*}
with $\zeta(x)$ the Riemann zeta function. 

In contrast, $\kappa_{T}/\kappa_{0}$ tends asymptotically to the cons\-tant $(1+\widetilde{m})^{-1/2}$ for $d=1$, while it goes to zero for $d<1$ as can be seen from Fig.\ref{fig:compressibility} for $d=1/2$. In this latter case, though the system behave qualitatively as standard matter, the effects due to pair creation can be noted quantitatively from the departure to the case when no pair creation is considered (thin-dashed lines).

The change of the temperature dependence of $\kappa_{T}$ that occurs at $T_{\kappa}^{*}$ for $d>1$, marks a drastic change in the qualitative behavior of the system driven by pair creation, suggesting a possible transition from a normal phase, to a phase in which the system becomes arbitrarily compressible with temperature. 

For the three-dimensional case we have that the minimum of $\kappa_{T}$ occurs at approximately at $T_{\kappa}^{*}=0.34732$ for $\widetilde{m}=0.01$, $T_{\kappa}^{*}=0.32018$ for $\widetilde{m}=1$, and $T_{\kappa}^{*}=0.07465$ for $\widetilde{m}=100$. 

At particle densities of the order of the nuclear matter 0.122 fm$^{-3}$, the corresponding Fermi energy is approximately $480.618$ MeV in the limit $\widetilde{m}\ll1$. With these values we estimate $k_{B}T_{\kappa}^{*}\simeq 166.928$ MeV for $\widetilde{m}=0.01$. This value is of the order of the expected crossover temperature to the quark-gluon plasma \cite{Martinez2013}, which from QCD calculations is expected to be 173$\pm15$ MeV for massless
quarks \cite{Karsch2002}, while QCD lattice calculations with non-zero quark masses ($m_{u}=2.3\pm0.5$ MeV, $m_{d}=4.8^{+0.7}_{-0.3}$ MeV \cite{BeringerPRD2012}) give crossover temperatures between 150 and 200 MeV \cite{Martinez2013}. Though for temperatures above $T_{\kappa}^{*}$ the ultrarelativistic ideal gas provides a good description of the quark-gluon plasma, for $T<T_{\kappa}^{*}$ the strong interactions become relevant in the thermodynamics of the system, interactions that our oversimplified approach neglects.  

\subsection{The specific heat at constant volume $C_{V}$}

The specific heat at constant volume is expressed in terms of the $\Lambda_{E_{k}}$'s as
\begin{widetext}
\begin{equation}\label{SpecificHeat}
C_{V}=\frac{R_{d}V_{d}}{k_{B}T^{2}}\int_{0}^{\infty}dk\, k^{d-1} E_{k}^{2}\left[\Lambda_{E_{k}}+\overline{\Lambda}_{E_{k}}\right]-
\frac{V_{d}}{\kappa_{T}T}\left[\frac{R_{d}}{n_{0}k_{B}T}\int_{0}^{\infty}dk\, k^{d-1}E_{k}\left[\Lambda_{E_{k}}+\overline{\Lambda}_{E_{k}}\right]\right]^{2}.
\end{equation}
\end{widetext}

\begin{figure}[h]
  \centering
   \includegraphics[width=0.7\columnwidth]{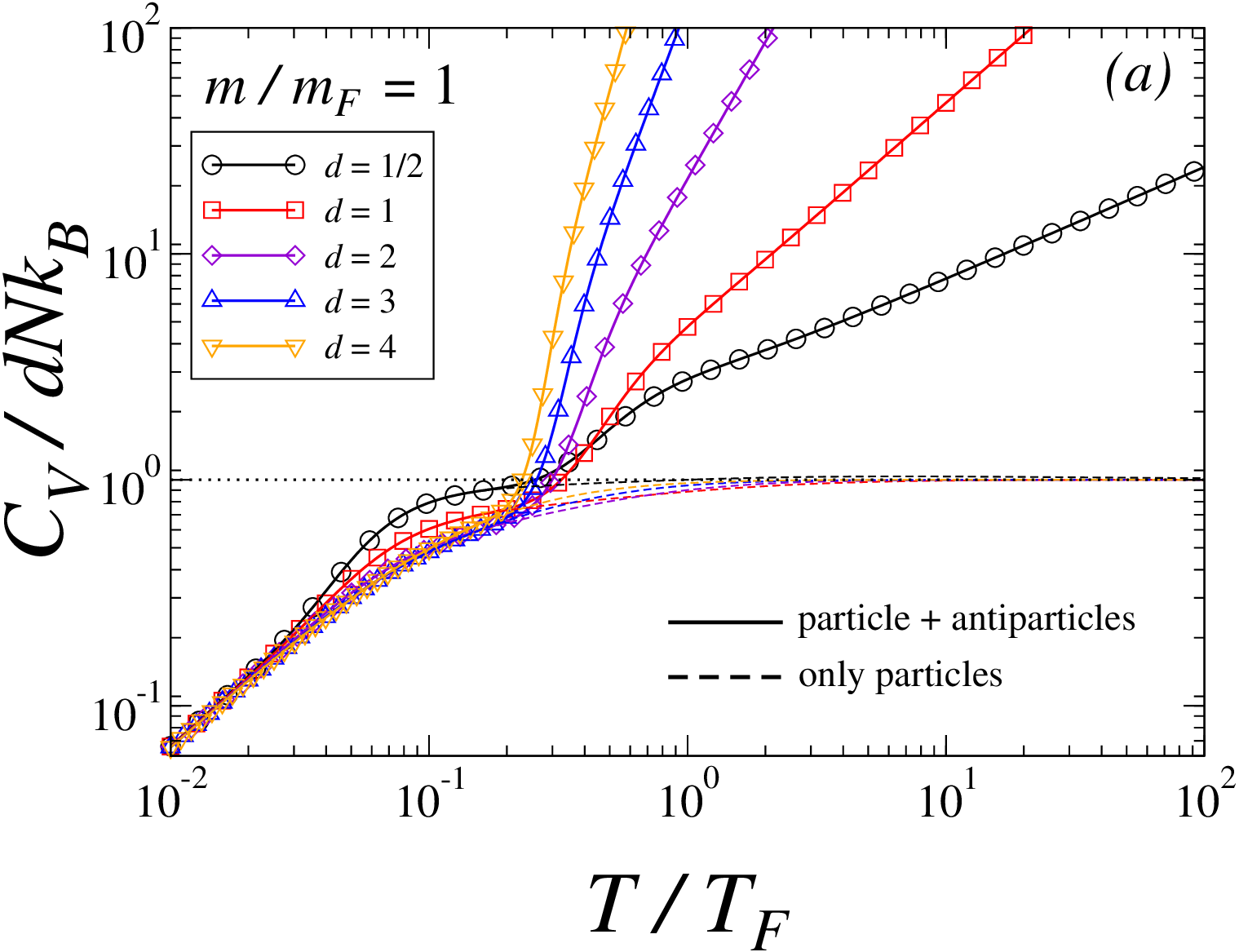}
   \includegraphics[width=0.7\columnwidth]{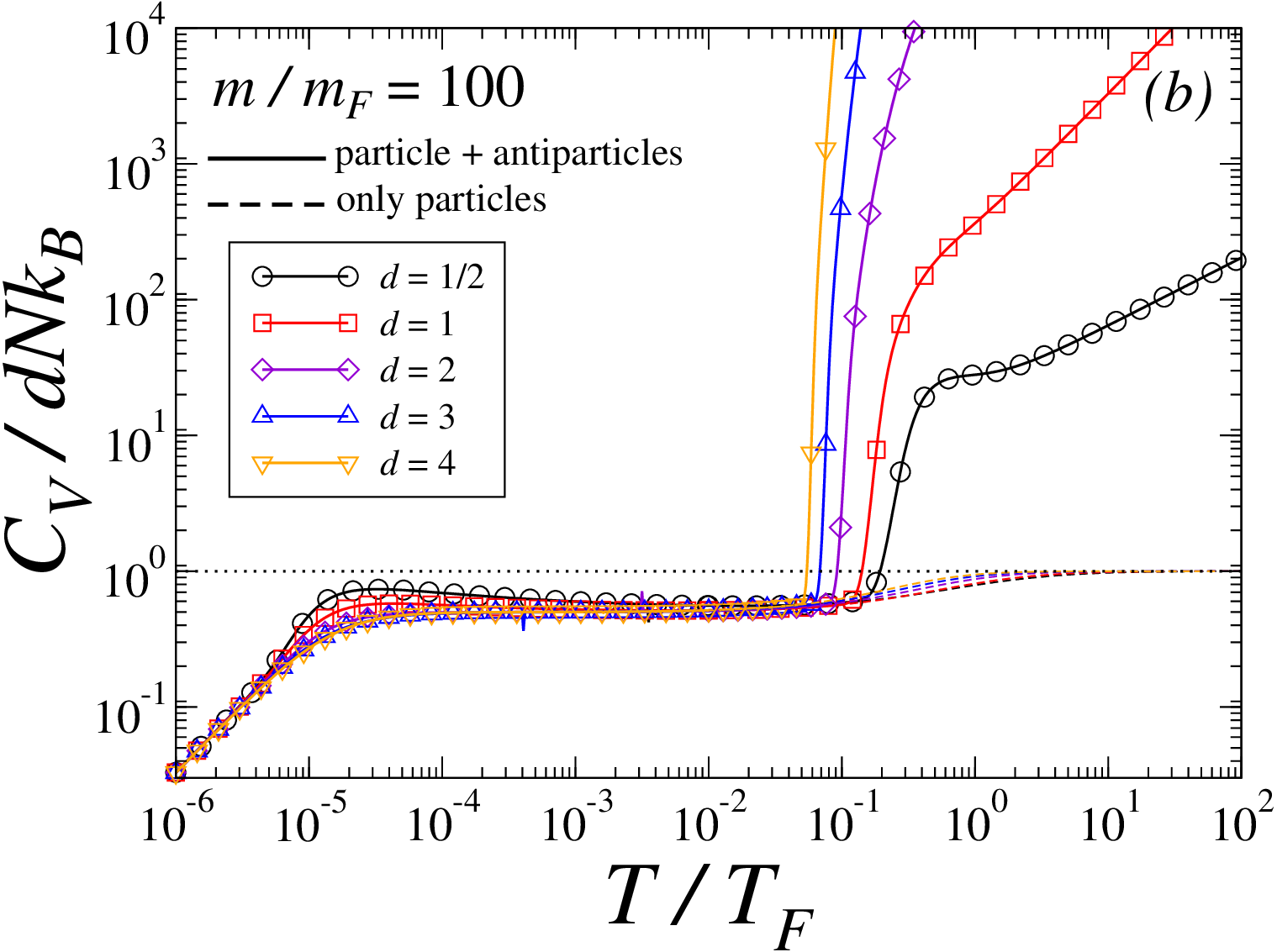}
  \caption{(Color online) Dimensionless specific heat at constant volume as function of the dimensionless temperature $T/T_{F}$ for dimension 1/2 (circles), 1 (squares), 2 (diamonds), 3 (up-triangles) and 4 (down-triangles) and mass ratio $\widetilde{m}=1$ [panel (a)], $\widetilde{m}=100$ [panel (b)]. Thin-dashed lines correspond to the cases for which pair production is neglected.}
\label{fig:SpecificHeat}
\end{figure}

As shown in Fig. \ref{fig:SpecificHeat}, the low temperature behavior is given by the well known linear dependence, with the pre\-factor $d\pi^{2}(1+\widetilde{m}^{2})/3$ which comes only from the Fermi-Dirac statistics of the particles and the dimensionality of the system. 

In the same range of temperatures where a local minimum in $\kappa_{T}$ is found, the specific heat changes its linear dependence characteristic of the low temperature regime, to the temperature dependence $T^{d}$ as is shown in Fig. \ref{fig:SpecificHeat}. For $\widetilde{m}\ll1$ (not shown in Fig. \ref{fig:SpecificHeat}) the transition is smooth and becomes more marked as the mass ratio increases, as is contrasted in panels (a) and (b), where a plateau appears before the power-law growth.   This behavior differs from the case for which only particles are considered (thin-dashed lines) which reaches the Dulong-Pettit limit $C_{V}/dNk_{B}=1$. 
The exact result $C_{V}/N_{0}k_{B}=\left(\pi^{2}/3\right)(T/T_{F})$ is found in the limit $\widetilde{m}\rightarrow0$ for $d=1$. 

In Ref. \cite{BlasPRE1999} the authors considered the relativistic Bose and Fermi gases, at low temperatures, they rightly neglected the antiparticles, and concluded  that in two dimensions both gases are thermodynamically inequivalent, in contrast to the non-relativistic case in which they do, however it seems they missed that both gases are thermodynamically equivalent in one dimension in the ultrarelativistic limit. In fact, it is known that the condition for the equivalence between the two quantum gases consists of the constancy of the single-particle density of states $g(E)$. In the exact relativistic case, is the finite rest mass of the particle what avoids such possibility, since there is no value of $d$ which makes the density of states $g(E)=E(E^{2}-m^{2}c^{4})^{d/2-1}\theta(E-mc^{2})$, $\theta(x)$ being the Heaviside step function, to be a constant as occurs in the non-relativistic and ultrarelativistic cases, where $g(E)\propto E^{d/2-1}$ and $\propto E^{d-1}$, respectively. 

\section{\label{IV}Conclusions and final remarks}
We have studied the effects of the system dimensionality and quantum-relativity on the thermodynamics of an ideal Fermi gas. 
The temperature dependence of the chemical potential is determined by the system dimensionality and by the particles rest mass. We recovered the unusual low temperature dependence of $\mu(T)$ for $d<2$ \cite{GretherEPJD2003} in the non-relativistic limit $m\gg m_{F}$. For arbitrary values of the rest mass, the nonmonotonic behavior of $\mu$ in the low temperature regime appears if $d<2-(1+\widetilde{m}^{2})^{-1}$, which includes the ultrarelativistic case for $\widetilde{m}\ll1$. 
Singularly, for dimensions smaller than one, $\mu$ increases monotonically with $T.$ This peculiar behavior occurs since for low dimensional systems, the creation of particle-antiparticle pairs occurs at a so low rate that the initial number of fermions dominates the thermodynamic behavior of the system. This argument is supported from the temperature dependence of $\kappa_{T}$ which vanishes as $T\rightarrow\infty,$ just as in the case when only particles are considered (dashed lines in Fig.\ref{fig:ChemPot}). The temperature dependence of $\mu$ for high temperatures described in Fig.\ref{fig:ChemPot}(a) is also observed in the relativistic Bose gas with pair production \cite{FrotaPRA1989} for $d>1$, with the remarkable difference that for the Bose gas, $\vert\mu_{B}\vert\le mc^{2}$, and therefore the chemical potential vanishes as $T\rightarrow\infty$ even for $0<d<1$. Except in the case $d=1$ for which we have $\mu_{B}=0$ for all $T$ where $\mu_{B}$ is the chemical potential of the Bose gas. 
  
The effects of pair production are exhibited  in the thermodynamical susceptibilities as a change in their temperature dependence that appears at some tenths of the Fermi temperature (as shown in Figs. \ref{fig:compressibility} and \ref{fig:SpecificHeat}) corresponding to the temperature range at which the pair production becomes significantly important. Both susceptibilities start growing without limit as a power law of $T$ after this crossover. The temperature that points out the crossover $T_{\kappa}^{*}$ could be determined from the apparent local minimum exhibited in the isothermal compressibility.

Interestingly for three dimensions and at particle densities of the size of the nuclear densities, the crossover occurs at approximately 167 MeV for a rest mass close to 4.86 MeV. This temperature is close to the quark-gluon plasma critical temperature $\lambda_{QCD}$ expected to occur for light quark masses. Above such a temperature the strong interaction among quarks can be neglected leading to the ideal situation described in this paper.     

Since our calculations exhibits the exact thermal behavior of the relativistic IFG, evidently can describe systems beyond the standard complete-degeneracy approximation ($T=0$), generally used in situations where a disparate difference between the system's and Fermi temperature exists, as in the case of white dwarf stars, where $T/T_{F}\simeq10^{-2}$. Since the effects of dimensionality on the thermodynamical susceptibilities at low temperatures are not negligible, it would be desirable to extend the analysis presented in Ref. \cite{ChavanisPRD2007} to incorporate finite temperature calculations, particularly for $d<3$, where the susceptibilities exhibit an anomalous behavior. By doing so, an study of the response of white dwarf stars to thermal and pressure fluctuations could plausibly establish a lower bound on the anthropic dimension of the Universe. Thus, in addition with the upper bound ($d<4$) for the anthropic dimension of the Universe given by Chavanis \cite{ChavanisPRD2007}, it would be plausibly justified the three spatial dimensions of the observed Universe.

\section{Acknowledgments}
The authors acknowledge financial support from DGAPA-UNAM grant PAPIIT-IN111070.
\section{Appendix: The zero temperature relativistic IFG}

The zero point energy per particle, $u_{0}=U_{0}/N_{0},$ 
can be written in terms of the Gaussian or ordinary hypergeometric function $_{2}F_{1}(a_{1},a_{2};b_{1};z)$ \cite{Abramowitz} as
\begin{align}
u_{0}=& mc^{2}\, _{2}F_{1}\left[-1/2,d/2;1+d/2;-\widetilde{m}^{-2}\right].
\end{align}
that reduces to elementary functions for integer values of $d$.
For $\widetilde{m}\gg1$ we have
\begin{equation}
u_{0}=mc^{2}+\frac{d/2}{1+d/2}E_{F}^{NR}-\frac{d/2}{2+d/2}\frac{\left(E_{F}^{NR}\right)^{2}}{mc^{2}}+\ldots
\end{equation}
In the opposite limit, $\widetilde{m}\ll1,$ we can write 
\begin{equation}
u_{0}=\frac{d}{d+1}E_{F}^{UR}\left[1+\frac{1}{2}\frac{d+1}{d-1}\widetilde{m}^{2}-\frac{1}{8}\frac{d+1}{d-3}\widetilde{m}^{4}+\ldots\right]
\end{equation}
for $d\neq1,\, 3,\, 5\ldots$. 

For $d=3$ and 1 we have, respectively 
\begin{align}
u_{0}&=\frac{3}{4}\, E_{F}^{UR}(1+\widetilde{m}^{2})^{1/2}\left[1+\frac{\widetilde{m}^{2}}{2}\left(1-\frac{\widetilde{m}^{2}\arcsinh(\widetilde{m}^{-1})}{(1+\widetilde{m}^{2})^{1/2}}\right)\right]\\
u_{0}&=\frac{1}{2}\, E_{F}^{UR}\left[(1+\widetilde{m}^{2})^{1/2}+\widetilde{m}^{2}\arcsinh(\widetilde{m}^{-1})\right],
\end{align}
In the $\widetilde{m}\ll1$ limit, last expressions can approximated by
\begin{align}
u_{0}&=\frac{3}{4}\, E_{F}^{UR}\left[1+\widetilde{m}^{2}+\frac{1}{8}\widetilde{m}^{4}\ln\left(\widetilde{m}\right)+\ldots\right]\\
u_{0}&=\frac{1}{2}\, E_{F}^{UR}\left[1-\widetilde{m}^{2}\ln\widetilde{m}+\frac{1}{8}\widetilde{m}^{4}+\ldots\right].
\end{align}

For the zero point pressure $P_{0}$ we have
\begin{align}
P_{0}/n_{0}=& m_{F}c^{2}\, \sqrt{1+\widetilde{m}^{2}} - u_{0}
\end{align}
which for $\widetilde{m}\gg1$ last expression reduces to
\begin{equation}
P_{0}/n_{0}=\frac{2}{d+2}E_{F}^{NR}+\frac{d/2}{2+d/2}\frac{\left(E_{F}^{NR}\right)^{2}}{mc^{2}}+\ldots
\end{equation}
where the first term corresponds to the well known non-relativistic case. In the opposite limit
\begin{equation}
P_{0}/n_{0}=\frac{1}{d+1}E_{F}^{UR}-\frac{1}{2(d-1)}E_{F}^{UR}\widetilde{m}^{2} +\frac{d}{8(d-1)}E_{F}^{UR}\widetilde{m}^{4}\ldots
\end{equation}
the first term corresponds to the well known result in the ultrarelativistic case, the next terms are valid always that $d$ is not an odd integer.
For $d=3,1$ we have respectively
\begin{align}
P_{0}/n_{0}=&\frac{1}{4}E_{F}^{UR}-\frac{1}{4}E_{F}^{UR}\widetilde{m}^{2} +\frac{3}{2^{5}}E_{F}^{UR}\widetilde{m}^{4}\ln\widetilde{m}\ldots\\
P_{0}/n_{0}=&\frac{1}{2}E_{F}^{UR}+\frac{1}{2}E_{F}^{UR}\widetilde{m}^{2}\left[1-\ln\widetilde{m}\right]-\frac{1}{2^{4}}E_{F}^{UR}\widetilde{m}^{4}+\ldots
\end{align}

The inverse of the isothermal compressibility $\kappa_{T}=-\left(1/V_{d}\right)\left(\partial V_{d}/\partial P_{0}\right)_{T}$ is given by
\begin{equation}
\kappa_{0}^{-1}=\frac{n_{0}}{d}\frac{m_{F}c^{2}}{\sqrt{1+\widetilde{m}^{2}}}
\end{equation}
in the limit of $\widetilde{m}\gg1$ we have that
\begin{equation}
\kappa_{0} \simeq \kappa_{0}^{NR}\left[1+\frac{1}{2\widetilde{m}^{2}}\right]
\end{equation}
where $\kappa_{0}^{NR}=d/[(d+2)P_{0}]$ is the NR isothermal compressibility, which reduces to the well known result $\kappa_{0}=(3/5)P_{0}^{-1}$ for $d=3$, and  
\begin{equation}\label{CompresibilityUR}
\kappa_{0} \simeq \kappa_{0}^{UR}\left[1+\frac{1}{2}\widetilde{m}^{2}\right]
\end{equation}
in the $\widetilde{m}\ll1$ case, where $\kappa_{0}^{UR}=d/[(d+1)P_{0}]$. These results show that the gas is more compressible than in their respective limits $\widetilde{m}\rightarrow\infty$ and $\widetilde{m}=0$.

\end{document}